\documentclass{midl} % Include author names
\usepackage{mwe} % to get dummy images
\usepackage{verbatim}

\jmlrproceedings{MIDL}{Medical Imaging with Deep Learning}
\jmlrpages{}
\jmlryear{2021}

\jmlrworkshop{Short Paper -- MIDL 2022 submission}
\jmlrvolume{-- Under Review}
\editors{Under Review for MIDL 2022}

\title[Short Title]{Development of an algorithm for medical image segmentation of bone tissue in interaction with metallic implants}

\midlauthor{\Name{Fernando {García Torres}\nametag{$^{1,2}$}} \Email{fernando.garcia@ibv.org}\\
\addr $^{1}$ Instituto de Biomecánica de València (IBV) \\
\addr $^{2}$ Universitat Politècnica de València (UPV) \AND
\Name{Carmen {Mínguez Porter}\nametag{$^{2}$}} \Email{carminpo@etsii.upv.es}\\
\Name{Julia {Tomás Chenoll}\midljointauthortext{Contributed equally}\nametag{$^{1,2}$}} \Email{julia.tomas@ibv.org}\\
\Name{Sofía {Iranzo Egea}\midlotherjointauthor\nametag{$^{1,2}$}} \Email{sofia.iranzo@ibv.org}\\
\Name{{Juan Manuel} {Belda Lois}\nametag{$^{1,2}$}} \Email{juanma.belda@ibv.org}\\
}

\begin{document}

\maketitle

\begin{abstract}
This preliminary study focuses on the development of a medical image segmentation algorithm based on artificial intelligence for calculating bone growth in contact with metallic implants. %as a result of the problem of estimating the growth of new bone tissue due to artifacts. %the presence of various types of distortions and errors, known as artifacts.
Two databases consisting of computerized microtomography images have been used throughout this work: 100 images for training and 196 images for testing. Both bone and implant tissue were manually segmented in the training data set. The type of network constructed follows the U-Net architecture, a convolutional neural network explicitly used for medical image segmentation. 
In terms of network accuracy, the model reached around 98\%. Once the prediction was obtained from the new data set (test set), the total number of pixels belonging to bone tissue was calculated. This volume is around 15\% of the volume estimated by conventional techniques, which are usually overestimated.
This method has shown its good performance and results, although it has a wide margin for improvement, modifying various parameters of the networks or using larger databases to improve training.

\end{abstract}

\begin{keywords}
Bone tissue segmentation, medical imaging, deep learning, CNN, U-Net
\end{keywords}

\section{Introduction}

One of the most widely used non-destructive medical imaging techniques is computed tomography (CT). Computed microtomography is a CT variant using micrometer resolutions. One of the uses of these techniques is to evaluate the performance of surgical implants in the body. Usually, the images obtained present various artifacts such as noise, low contrast or blurred areas. Some techniques are often applied to reduce these effects  \cite{boas_ct_2012}. 
Regarding evaluating the interaction of body tissues with metallic implants, semi-automated methods of commercial medical imaging software currently overestimate the amount of tissue due to image defects and artifacts \cite{ripley_3d_2017}. 
In addition, the progress of machine learning and artificial intelligence has led to the development of new techniques that have been shown to play an important role in medical image processing and image segmentation \cite{zhang_ct_2018}. The purpose of this work is to assess the feasibility of using convolutional neural networks (CNN) to improve the results obtained by the current \textmu CT image reconstruction and segmentation software, Mimics Innovation Suite (Materialise, Belgium), in CT images of cylindrical metal implants inserted in rabbit distal femoral condyles. This program performs a 3D reconstruction from 2D images and perform various calculations, such as the volume of regenerated bone around a prosthesis, but it introduces overestimates due to the artifacts caused by metal implants.

\section{Methods}
%\subsection{Manual image segmentation}

ImageJ software was used for the segmentation of the image set. The process consisted of: image preprocessing to improve viewing conditions, manual segmentation, review and correction. The images were segmented by three cases: image background (pixel value 0), bone (pixel value 1) and implant (pixel value 2).

%\subsection{Dataset configuration and network architecture}
The dataset consisted of 100 images in grayscale and size 2016x2016 pixels, that were downscaled to 512x512 pixels. Bone and implant masks were transformed from grayscale to binary images. The dataset was divided into 95 training and five validation images. A new database of 196 images was used as a testing dataset, belonging to a complete new sample. From the predictions made on this second database, the calculation of bone volume was performed and later compared with the results obtained by Mimics.

The network designed with 2.6.0 version of Keras \cite{chollet_keras_2015} followed the structure of a U-Net, which consists of a contraction path with 2D convolutional layers and max-pooling filters, and an expansion path, with transposed convolution layers. Two convolutions were carried out for each step, increasing the depth of the images and going from 1 channel to 64. After each convolution, a ReLU activation function was applied. A max-pooling process was applied after, reducing the image size by half. These steps were repeated until the last step (without max-pooling), where the expansion path began to obtain the image in its original size. The last layer of the network had sigmoid activation. The loss function used was categorical cross-entropy and the optimizer selected was Adam.

%\subsection{Volume comparison}
Volume calculation was estimated by the number of pixels belonging to the bone class of the test dataset predictions. This parameter was compared with the pixels calculated in the Mimics software (\ref{eq:1})   where \(V_C\) and \(pixels_C\) are the parameters calculated from the CNN prediction and \(V_M\) and \(pixels_M\) from Mimics software.

\begin{equation}
    \frac{V_C (mm^3)}{pixels_C} = \frac{V_M (mm^3)}{pixels_M} \label{eq:1}
\end{equation}

\section{Results}
The network trained with $512 \times 512$ images for 50 epochs achieved an accuracy of 98\% (\figureref{fig:accuracy}) and reached a negligible level of loss. The total pixel count predicted was $4,154,096$, and the total pixel count in Mimics $23,546,219$ pixels, which is equivalent to $365.03 mm^3$. Therefore, the volume of bone provided by the network is $64.40 mm^3$, approximately $17.64\%$ the volume calculated with Mimics.

\section{Discussion}

\begin{figure}[htbp]
 % Caption and label go in the first argument and the figure contents
 % go in the second argument  
%\floatconts
%  
  {\caption{Results obtained from the CNN.}}
  \begin{subfigure}[model's accuracy]
  {\includegraphics[width=0.4\linewidth]{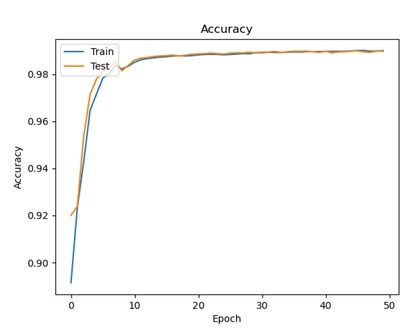}\label{fig:accuracy}}
  \end{subfigure}
  \begin{subfigure}[comparison of performances]
  {\includegraphics[width=0.4\linewidth]{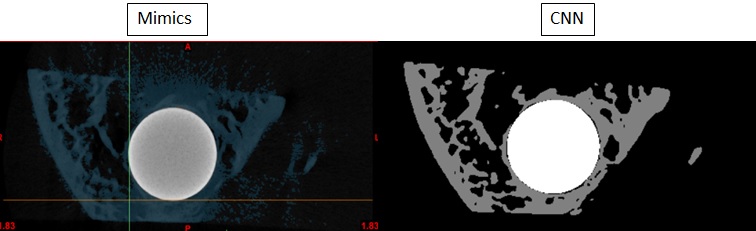}\label{fig:implantes}}
  \end{subfigure}
\end{figure}

Despite being a small dataset of CT, the results of the CNN achieved acceptable performance. The CNN reduced the estimation of the bone volume considerably compared with the estimation of Mimics software. On a qualitative level, the segmentation performed by Mimics seemed to be noisier, more blurred and with lower contrast (\figureref{fig:implantes}). It should be noted that the model cannot remove the effect of metallic artifacts completely, as it did not have images with significant artifacts, which may lead to an overestimation of the bone volume. A more significant number of training images with a larger number of artifact typologies would be necessary to robust the network. %to filter out noise in a larger number of cases.

\section{Conclusions}
Despite the limited number of train images, the network performs acceptably, eliminating much of the noise produced by the radiation on the metal implant and overestimating the bone volume to a lesser extent than commercial software.

% Acknowledgments---Will not appear in anonymized version
%\midlacknowledgments{We thank a bunch of people.}

%\bibliographystyle{wileyj}
\bibliography{Articulo}

\end{document}